\documentstyle[12pt]{article}
\newcommand{\be}{\begin{equation}}
\newcommand{\ee}{\end{equation}}
\newcommand{\bea}{\begin{eqnarray}}
\newcommand{\eea}{\end{eqnarray}}

\newcommand{\p}[1]{(\ref{#1})}

\begin{document}
\begin{titlepage}
%\begin{flushright}
%hep-th/0606053\\
%June 2006
%\end{flushright}
\vspace*{1.2cm}
\vskip 0.9truecm

\begin{center}
{\Large\bf Massless higher spin $D=4$ superparticle with both}

\vspace{0.3cm}
{\Large\bf $N=1$ supersymmetry and its bosonic counterpart}

\vspace{1.5cm}
\renewcommand{\thefootnote}{\star}

{\large\bf Sergey Fedoruk$^1$}\footnote{On leave from Ukrainian
Engineering-Pedagogical Academy, Kharkov, Ukraine},\,\,\,
{\large\bf Evgeny Ivanov$^1$},\,\,\,
\renewcommand{\thefootnote}{\dag}
{\large\bf Jerzy Lukierski$^2$}\footnote{Supported by KBN grant 1
P03B 01828}

\vskip 1cm

 \ $^1${\it Bogoliubov Laboratory of
Theoretical Physics, JINR, \\ 141980 Dubna, Moscow Region, Russian Federation}, \\
{\tt fedoruk,eivanov@theor.jinr.ru}

\vskip 0.5cm

\ ${}^2${\it Institute for Theoretical Physics, University of Wroc{\l}aw} \\
{\it pl. Maxa Borna 9, 50-204 Wroc{\l}aw, Poland},\\
{\tt lukier@ift.wroc.pl}

\end{center}
\vspace{0.2cm}
\vskip 0.6truecm  \nopagebreak

\begin{abstract}
\noindent We show that the massless higher spin four-dimensional
particle model with bosonic counterpart of $N=1$ supersymmetry
respects $SU(3,2)$ invariance. We extend this particle model to a
superparticle possessing $SU(3,2|1)$ supersymmetry which is a
closure of standard four-dimensional $N=1$ superconformal symmetry
$SU(2,2|1)$ and its bosonic $SU(3,2)$ counterpart. The new massless
higher spin $D=4$ superparticle model describes trajectories in
Minkowski superspace $(x^\mu, \theta^\alpha,
\bar\theta^{\dot\alpha})$ extended by the commuting Weyl spinor
$\zeta^\alpha$, $\bar\zeta^{\dot\alpha}=(\zeta^\alpha)^\ast$. We
find the relevant phase space constraints and quantize the model. As
a result of quantization we obtain the superwave function which
describes an infinite sequence of four-dimensional massless chiral
superfields with arbitrary external helicity indices.
\end{abstract}

\newpage

\end{titlepage}
\setcounter{footnote}{0}

\section{Introduction}

A concise approach to the field--theoretic description of
higher spin (HS) multiplets (see {\it e.~g.}~\cite{FrVas}--\cite{Sor})
is based on extending the Minkowski space by additional bosonic coordinates.
Two such extensions are presently known:

\smallskip
{\it i)} We can add to four Minkowski coordinates $x^\mu
={\textstyle\frac{1}{2}} \sigma^\mu_{\alpha\dot\beta}
x^{\dot\beta\alpha}$ six \underline{tensorial} ones $z^{[\mu\nu]}
={\textstyle\frac{1}{2}} \sigma^{[\mu\nu]}_{\alpha\beta}
z^{\alpha\beta}+ {\textstyle\frac{1}{2}}
\sigma^{[\mu\nu]}_{\dot\alpha\dot\beta} \bar
z^{\dot\alpha\dot\beta}$ and obtain the tensorially extended
space--time ${\cal M}^{4;6}_T$~ \cite{Frons}--\cite{BPST}. Ten
coordinates  $X^{M}=(x^\mu, z^{[\mu\nu]})$ can be combined into a
symmetric second rank $Sp(4)$ (or $GL(4)$) spinor $X^{(ab)}$
($a,b=1,...,4$). The translations in ${\cal M}^{4;6}_T$ are
generated by the operators $P_{ab} =(P_\mu, Z_{\mu\nu})$ which
naturally occur in of $N=1$, $D=4$ superalgebra extended by
tensorial charges \cite{FerPor,RS}.

The tensorial particle in the mixed space-time--twistor formulation is described
by the following action \cite{BandLuk,BandLukSor}
\begin{equation}\label{act-sp-mix}
S = \int d\tau \,
\Lambda_a \Lambda_b \dot X^{ab} \,,
\end{equation}
where $\Lambda_a$ is a real $\overline{O(3,2)}\simeq Sp(4)$ spinor.
The space of its quantum states contains an infinite number of
four-dimensional massless states
with all integer and half-integer helicities. They are
described by a wave function given on the extended space
${\cal M}^{4;6}_T =(X^{ab}, \Lambda_a)\,$. Its Fourier transform with respect to
$\Lambda_a$ obeys
the unfolded form of the conformal HS equations \cite{Vas}.
In such a framework, the role of Lorentz group is played by the group $GL(4)$ and
the conformal symmetry
$SU(2,2)\simeq \overline{SO(4,2)}$ is extended to $Sp(8)$ which
is nonlinearly realized in ${\cal M}^{4;6}_T$.

Using in~\p{act-sp-mix} the generalized Penrose incidence relation
\begin{equation}\label{inc-sp}
M^a = X^{ab} \Lambda_b\,,
\end{equation}
one obtains, modulo boundary terms,
the purely twistorial form of the HS particle action
\begin{equation}\label{act-sp-mix1}
S = 2\int d\tau \,
\Lambda_a \dot M^a  =
 \int d\tau \,
Y_A \dot Y^A \,.
\end{equation}
Here $Y_A =(\Lambda_a, M^b )$, $A=1,...,8$, is the fundamental $Sp(8)\,$ spinor.
Since in the purely twistorial formulation the generalized conformal symmetry $Sp(8)$ acts
as linear transformations of $Y_A$, this formulation provides
the most transparent symmetry structure.

An extension to the tensorial $N=1$ superparticle was given in \cite{BandLuk,BandLukSor}.
As a result of quantization of such a system, the
superfield HS equations were derived, which are covariant
under the generalized $N=1$ superconformal group $OSp(1\vert 8)\,$.

\smallskip
{\it ii)}
Another way of deriving
infinite HS multiplets was recently proposed in \cite{FedLuk,FedIv}. It
consists in adding, to the four space--time coordinates $x^{\alpha\dot\beta}\,$,
commuting \underline{spinorial} variables $\zeta^\alpha$,
$\bar\zeta^{\dot\alpha}= (\overline{\zeta^\alpha})\,$. In this way one obtains
the spinorial extension of the  Minkowski space, ${\cal M}^{4;4}_S\,$.
As shown in~\cite{FedLuk}, quantization of
bosonic counterpart of the $N=1$, $D=4$ Brink--Schwarz model~\cite{BS},
\begin{equation}\label{Lagr}
S = \int d\tau\, {\textstyle\frac{1}{2e}}\,
\omega^{\dot\alpha\alpha}\omega_{\alpha\dot\alpha}
\end{equation}
where
\begin{equation}\label{om}
\omega^{\dot\alpha\alpha} = \dot x^{\dot\alpha\alpha}-i
{\bar\zeta}{}^{\dot\alpha}\dot{\zeta}{}^{\alpha}+i
\dot{\bar\zeta}{}^{\dot\alpha}{\zeta}^{\alpha}\,,
\end{equation}
yields the wave function
which is given on ${\cal M}^{4;4}_S\,$ and accommodates,
in its $\zeta^\alpha$,
$\bar\zeta^{\dot\alpha}$ expansion, an infinite set of massless states with
arbitrary (integer and half--integer)
helicities. The model ~(\ref{Lagr}) is invariant under the bosonic
counterpart of $N=1$, $D=4$ supersymmetry.

\smallskip
First aim of the present paper is to study in
detail the symmetry structure
of the model~(\ref{Lagr}).
In Sect.~2 we pass to the purely twistorial description of the model and
demonstrate that it exhibits the
$SU(3,2) \supset SU(2,2)$ symmetry
in analogy with the $Sp(8) \supset SU(2,2) $ symmetry
of the tensorial particle.
It appears that $SU(3,2)$ is the
bosonic counterpart of $N=1$, $D=4$ superconformal symmetry $SU(2,2\vert 1)\,$.
Our second aim is to construct $N=1$ superextension
of~(\ref{Lagr}).
In Sect.~3 we supersymmetrize the model~(\ref{Lagr}) by passing to the superparticle moving
in the spinorially extended Minkowski superspace $(x^{\alpha\dot\alpha} ,\zeta^\alpha,
\bar\zeta^{\dot\alpha}, \theta^\alpha,
\bar\theta^{\dot\alpha})$, with $\theta^\alpha$ being the Grassmann spinor coordinate.
We show that the conformal supersymmetry of such a model is $SU(3,2|1)$.
The relations between various (super)conformal groups are
illustrated
on the diagram\\
\begin{picture}(150,30)
\put(50,24){$SU(2,2)$}
\put(105,24){$SU(2,2|1)$}
\put(50,4){$SU(3,2)$}
\put(105,4){$SU(3,2|1)$\,.}
\put(70,26){\vector(1,0){30}}
\put(70,6){\vector(1,0){30}}
\put(55,20){\vector(0,-1){10}}
\put(115,20){\vector(0,-1){10}}
\end{picture}\\
Here the vertical arrows indicate the ``bosonic supersymmetrization'',
while the horizontal ones the $N=1$ supersymmetrization.
In Sect.~4 we quantize our model and obtain an infinite tower
of chiral $N=1$ supermultiplets with growing external helicities.
Sect.~5 contains summary and outlook.

\setcounter{equation}{0}
\section{HS particle with even SUSY and its $SU(3,2)$ invariance}

There are three (equivalent) formulations of
the HS particle model~(\ref{Lagr}).

{\it i) Pure space--time formulation.}
The action~(\ref{Lagr}) rewritten in the first order form is
\begin{equation}\label{act-bsusy}
S=\int d\tau \left( P_{\alpha\dot\alpha} \omega^{\dot\alpha\alpha} -
e P_{\alpha\dot\alpha}P^{\alpha\dot\alpha} \right)
\end{equation}
where $\omega^{\dot\alpha\alpha}$ is defined in~(\ref{om}).
Though the action~(\ref{act-bsusy}) looks like the Brink--Schwarz action,
its distinguishing feature is the commutativity of the spinor variables
$\zeta_\alpha, \bar\zeta_{\dot\alpha}\,$.

{\it ii) Mixed twistorial--space--time formulation.}
Using the standard expression for a light--like vector in terms of
the commuting spinor $\lambda_\alpha\,$,
\begin{equation}\label{re-p}
P_{\alpha\dot\alpha}=\lambda_\alpha \bar\lambda_{\dot\alpha}\,,
\end{equation}
we obtain an equivalent formulation of the same HS particle, with the action
\begin{equation}\label{act-bsusy-mix}
S=\int d\tau  \lambda_\alpha \bar\lambda_{\dot\alpha} \omega^{\dot\alpha\alpha} .
\end{equation}
The action~(\ref{act-bsusy-mix}) describes a bosonic counterpart of
the Ferber--Shirafuji superparticle~\cite{Fer,Shir}.
The spinor $\lambda_\alpha$ can be identified with half of the Penrose twistor~\cite{PenMac}.

{\it iii) Pure twistorial formulation.}
Space--time variables are eliminated by introducing the second half of
the twistor, the  complex spinor  $\mu^\alpha\,$, as well as an additional
complex scalar variable $\xi\,$. The bosonic counterpart of the supertwistorial
incidence relation~\cite{Fer} is
\begin{equation}\label{inc}
\bar \mu^{\dot\alpha} =x^{\dot\alpha\alpha}\lambda_{\alpha}
+ i\, \bar\zeta^{\dot\alpha} \xi \,, \qquad
\xi =\zeta^\alpha \lambda_\alpha \quad \mbox{and c.c.}\,.
\end{equation}
The scalar $\xi$ is the bosonic counterpart of
the odd component of the $SU(2,2|1)$ supertwistor.
After inserting the twistor transform~(\ref{inc}) into the action~(\ref{act-bsusy-mix}),
we obtain its twistorial form
\begin{equation}\label{act-twist}
S =\int d\tau \left( \lambda_\alpha\dot \mu^\alpha
+\bar\lambda_{\dot\alpha}
\dot{\bar \mu}^{\dot\alpha}
+i(\dot{\bar\xi}\xi - \bar\xi\dot\xi) -\Lambda U \right).
\end{equation}
The constraint added in~(\ref{act-twist}) involves the $SU(3,2)$
norm of the twistor $(\lambda_\alpha, {\bar \mu}^{\dot\alpha}, \xi)$
\begin{equation}\label{n-tw}
U \equiv i(\mu^\alpha\lambda_\alpha
-\bar\lambda_{\dot\alpha} {\bar \mu}^{\dot\alpha}) - 2\bar\xi\xi \approx 0\,.
\end{equation}
The vanishing of this norm follows from the hermitian property of
$x^{\alpha\dot\alpha}$. The Lagrange multiplier $\Lambda$ can be
regarded as $U(1)$ gauge field ensuring the invariance of the
action under local phase transformations of the involved fields,
and  $U$ as the corresponding $U(1)$ current.

It was recently shown \cite{FedIv} that, by enlarging the phase space,
one can unify the space--time and purely twistorial description.
In such a master model the actions~(\ref{act-bsusy}) and~(\ref{act-twist})
are recovered after fixing some local gauge invariances.

Let us now study the symmetries of~(\ref{act-bsusy}).
The even supertranslations read
\begin{equation}\label{bsusy-tran}
\delta x^{\dot\alpha\alpha} = i(\bar\epsilon^{\dot\alpha}\zeta^\alpha
-\bar\zeta^{\dot\alpha}\epsilon^\alpha) \, ,\quad \delta
\zeta^\alpha = \epsilon^\alpha \, ,\quad \delta
\bar\zeta^{\dot\alpha}= \bar\epsilon^{\dot\alpha}
\end{equation}
where $\epsilon^\alpha$ is a constant commuting spinor.
Remaining symmetries are transparent in the twistorial formulation~(\ref{act-twist})
(see {\it e. g.}~\cite{BC,Town}).
Below we list the generators of all such symmetries.

{}From the action~(\ref{act-twist}) we obtain the following
canonical Poisson brackets
\begin{equation}\label{cpb-tw}
[\mu^{\alpha} , \lambda_{\beta} ]_{{}_{\rm P}}=\delta^{\alpha}_{\beta} \,,
\qquad [\bar\mu^{\dot\alpha} , \bar\lambda_{\dot\beta} ]_{{}_{\rm P}}=
\delta^{\dot\alpha}_{\dot\beta}\,,  \qquad [\xi, \bar\xi ]_{{}_{\rm P}}=
{\textstyle\frac{i}{2}} \,.
\end{equation}
The finite--dimensional symmetries of the model~(\ref{act-twist}) are generated by
all the bilinear products
of phase variables~(\ref{cpb-tw})
which commute with the constraint~(\ref{n-tw}).
We obtain:\\
\noindent $\bullet$
The generators of $SU(2,2)$ transformations
\begin{equation}\label{P,K}
P_{\alpha\dot\alpha}=\lambda_\alpha \bar\lambda_{\dot\alpha}\,, \qquad K^{\dot\alpha\alpha}
=\bar \mu^{\dot\alpha} \mu^{\alpha}\,,
\end{equation}
\begin{equation}\label{L}
L_{\alpha\beta}=\lambda_{(\alpha}\mu_{\beta)} \,,\qquad \bar
L_{\dot\alpha\dot\beta}=\bar\lambda_{(\dot\alpha}
\bar \mu_{\dot\beta)} \,,
\end{equation}
\begin{equation}\label{D}
D= {\textstyle\frac{1}{2}}(\mu^\alpha\lambda_\alpha
+\bar\lambda_{\dot\alpha} {\bar \mu}^{\dot\alpha})\,;
\end{equation}
\noindent $\bullet$
The generators of the even SUSY translations and even superconformal boosts
\begin{equation}\label{R}
R_{\alpha}=-2i\,\bar\xi\lambda_{\alpha}\,, \qquad \bar
R_{\dot\alpha}=2i\,\xi\bar\lambda_{\dot\alpha}\,,
\end{equation}
\begin{equation}\label{C}
C^{\alpha}=-2i\,\xi\,\mu^{\alpha}\,, \qquad \bar
C^{\dot\alpha}=2i\,\bar\xi\,\bar \mu^{\dot\alpha}\,;
\end{equation}
\noindent $\bullet$
The generator of $U(1)$ transformations
\begin{equation}\label{A}
A = {\textstyle\frac{i}{2}}(\mu^\alpha\lambda_\alpha
-\bar\lambda_{\dot\alpha} {\bar \mu}^{\dot\alpha}) +4 \bar\xi\xi \,.
\end{equation}

The generators~(\ref{P,K})--(\ref{A}) form the oscillator representation of the
$SU(3,2)$ algebra. We present here
only the bosonic counterparts of the anticommutators of $SU(2,2|1)$:
$$
[R_{\alpha}, \bar R_{\dot\alpha} ]_{{}_{\rm P}}= -2i P_{\alpha\dot\alpha}\,,
\qquad [C^{\alpha}, \bar C^{\dot\alpha} ]_{{}_{\rm P}}= 2i K^{\dot\alpha\alpha}\,,
$$
$$
[R_{\alpha}, C^{\beta} ]_{{}_{\rm P}}= 2i L_{\alpha}{}^{\beta} + i (D-iA)
\delta_{\alpha}{}^{\beta}\,,
\qquad [\bar R_{\dot\alpha}, \bar C^{\dot\beta} ]_{{}_{\rm P}}=-2i
\bar L_{\dot\alpha}{}^{\dot\beta}
- i (D+iA) \delta_{\dot\alpha}{}^{\dot\beta}\,.
$$
Using \p{cpb-tw} and the explicit expressions~(\ref{P,K})--(\ref{A}), one can
find the transformations of the twistor variables
$(\lambda_{\alpha}, \bar\mu^{\dot\alpha}, \xi)$ and confirm that
these variables constitute the linearly transforming $SU(3,2)$
spinor,\footnote{Due to the interpretation of $SU(3,2)$ as extended
conformal group it will be also called $SU(3,2)$ twistor.}
with $U$ in \p{n-tw} as its invariant norm. For example,
the generators~(\ref{R}), (\ref{C}) produce the following
bosonic counterpart of the supersymmetry transformations
\begin{equation}\label{tr-tw}
\delta \lambda_{\alpha} = 2i \rho_{\alpha} \xi
\, ,\qquad \delta \bar\mu^{\dot\alpha} = 2i  \bar\epsilon^{\dot\alpha} \xi\,,
\qquad
\delta \xi = \epsilon^{\alpha}\lambda_{\alpha}
- \bar\rho_{\dot\alpha} \bar\mu^{\dot\alpha}\,,
\end{equation}
where $\epsilon^{\alpha}$ and $\rho_{\alpha}$ are parameters of the even SUSY and
the even superconformal boosts.

\setcounter{equation}{0}
\section{HS superparticle with even and standard SUSY}

\subsection{Action principle}

Superextension of the HS particle model~(\ref{act-bsusy}) can be obtained
via enlarging the space
${\cal M}^{4;4}_S$ by Grassmann Weyl spinors $\theta^\alpha$,
$\bar\theta^{\dot\alpha}= (\overline{\theta^\alpha})\,$.
The action of the HS superparticle with even SUSY is
the following superextension of~(\ref{act-bsusy})
\begin{equation}\label{act-bsusy-f}
S=\int d\tau \left( P_{\alpha\dot\alpha} W^{\dot\alpha\alpha} -
e P_{\alpha\dot\alpha}P^{\alpha\dot\alpha} \right)
\end{equation}
where
\begin{equation}\label{om-f}
W^{\dot\alpha\alpha} = \dot x^{\dot\alpha\alpha}-i
{\bar\theta}{}^{\dot\alpha}\dot{\theta}{}^{\alpha}+i
\dot{\bar\theta}{}^{\dot\alpha}{\theta}^{\alpha} -i
{\bar\zeta}{}^{\dot\alpha}\dot{\zeta}{}^{\alpha}+i
\dot{\bar\zeta}{}^{\dot\alpha}{\zeta}^{\alpha}\,.
\end{equation}
The action~(\ref{act-bsusy-f}) can be also treated as a HS generalization of
the $N=1$, $D=4$ Brink--Schwarz superparticle action \cite{BS} by means of
adding extra even spinor variables
$\zeta^\alpha$, $\bar\zeta^{\dot\alpha}$.

The supertwistorial formulation of the action~(\ref{act-bsusy-f})
is given in terms of two commuting Weyl spinors
$\lambda_\alpha$, $\bar \mu^{\dot\alpha}$ and two complex scalars $\xi$, $\chi\,$,
respectively commuting and anticommuting. Together they form a $SU(3,2|1)$ supertwistor.
The relations~(\ref{inc}) are generalized as follows:
\begin{equation}\label{inc-f}
\bar \mu^{\dot\alpha}
=x^{\dot\alpha\alpha}\lambda_{\alpha}  + i\,
\bar\theta^{\dot\alpha} \chi + i\, \bar\zeta^{\dot\alpha} \xi\,, \qquad
\chi =\theta^\alpha \lambda_\alpha \, ,\qquad
\xi =\zeta^\alpha \lambda_\alpha\,.
\end{equation}
{}For the hermitian $x^{\dot\alpha\alpha}$ they imply the
constraint on the $SU(3,2|1)$ norm
\begin{equation}\label{H-s}
{\cal U} \equiv i(\mu^\alpha\lambda_\alpha
-\bar\lambda_{\dot\alpha} {\bar \mu}^{\dot\alpha}) - 2\bar\xi\xi -  2\bar\chi\chi \approx 0\,.
\end{equation}

After performing the supertwistor transform~(\ref{inc-f}) and (\ref{re-p})
and adding the constraint \p{H-s}
with a Lagrange multiplier, the action~(\ref{act-bsusy-f}),
up to a boundary term, takes the form
\begin{equation}\label{act-twist-s}
S =\int d\tau \left[ \lambda_\alpha\dot \mu^\alpha
+\bar\lambda_{\dot\alpha}
\dot{\bar \mu}^{\dot\alpha} +i(\dot{\bar\xi}\xi - \bar\xi\dot\xi) +i(\dot{\bar\chi}\chi
- \bar\chi\dot\chi)-\Lambda {\cal U} \right].
\end{equation}
Like in the bosonic case, ${\cal U}$ in \p{act-twist-s} is the $U(1)$ current, while
$\Lambda(\tau)$
is the $U(1)$ gauge field.

\subsection{$SU(3,2|1)$ extension of superconformal symmetry}

The symmetries of our model are most
transparent in the twistorial formulation~(\ref{act-twist-s}).
The six--component supertwistor and its complex conjugate
span the complete phase space. The nonvanishing Poisson brackets are given by
eqs.~(\ref{cpb-tw}) and by
\begin{equation}\label{CPB-tw}
\{\chi, \bar\chi \}_{{}_{\rm P}}={\textstyle\frac{i}{2}} \,.
\end{equation}
The full set of the supersymmetry generators
preserving the constraint~(\ref{H-s}) consists of the generators~(\ref{P,K})--(\ref{A})
and new generators involving the odd scalar $\chi\,$. These are:

\noindent
$\bullet$ The generators of the standard
(odd) SUSY translations and odd superconformal boosts
\begin{equation}\label{Q}
Q_{\alpha}=2i\,\bar\chi\lambda_{\alpha}\,, \qquad \bar
Q_{\dot\alpha}=-2i\,\chi\bar\lambda_{\dot\alpha}\,,
\end{equation}
\begin{equation}\label{S}
S^{\alpha}=2i\,\chi\,\mu^{\alpha}\,, \qquad \bar
S^{\dot\alpha}=-2i\,\bar\chi\,\bar \mu^{\dot\alpha}\,;
\end{equation}
\noindent
$\bullet$
The generators of the additional odd symmetries
\begin{equation}\label{I}
I=2i\,\bar\xi\,\chi\,, \qquad \bar
I=-2i\,\xi\,\bar\chi\,;
\end{equation}
\noindent $\bullet$
The second $U(1)$ generator
\begin{equation}\label{A2}
\tilde{A} = {\textstyle\frac{i}{2}}(\mu^\alpha\lambda_\alpha
-\bar\lambda_{\dot\alpha} {\bar \mu}^{\dot\alpha}) -4 \bar\chi\chi\,.
\end{equation}
The set of the generators~(\ref{P,K})--(\ref{A}) and~(\ref{Q})--(\ref{A2})
form the $SU(3,2|\,1)$ algebra. We see therefore that the bilinear
form ${\cal U}$ in~(\ref{H-s}) is
the invariant norm in the graded $SU(3,2|\,1)$ supertwistor space.

Let us discuss some details of this superalgebra.

The generators~(\ref{Q}), (\ref{S}) satisfy the standard anticommutation relations
$$
\{Q_{\alpha}, \bar Q_{\dot\alpha} \}_{{}_{\rm P}}= 2i P_{\alpha\dot\alpha}\,,
\qquad \{S^{\alpha}, \bar S^{\dot\alpha} \}_{{}_{\rm P}}= 2i K^{\dot\alpha\alpha}\,,
$$
$$
\{Q_{\alpha}, S^{\beta} \}_{{}_{\rm P}}= -2i L_{\alpha}{}^{\beta} - i (D-i\tilde{A})
\delta_{\alpha}{}^{\beta}\,,
\qquad \{\bar Q_{\dot\alpha}, \bar S^{\dot\beta} \}_{{}_{\rm P}}
= -2i \bar L_{\dot\alpha}{}^{\dot\beta} - i (D+i\tilde{A}) \delta_{\dot\alpha}{}^{\dot\beta}
$$
and form, together with the $SU(2,2)$ generators~(\ref{P,K})--(\ref{D}) and
$U(1)$ generator~(\ref{A2}), $N=1$, $D=4$ superconformal algebra $SU(2,2|1)$.

The closure of the odd and even SUSY generators~(\ref{Q}), (\ref{S}) and~(\ref{R}), (\ref{C}),
$$
[Q_{\alpha}, C^{\beta} ]_{{}_{\rm P}}= -2i \bar I\delta_{\alpha}{}^{\beta}\,,
\qquad [R_{\alpha}, S^{\beta} ]_{{}_{\rm P}}= 2i I\delta_{\alpha}{}^{\beta}
\quad \mbox{and c.~c.}
$$
contains two additional odd generators~(\ref{I}). Together with the linear combination
\begin{equation}\label{U}
B = {\textstyle\frac{1}{2}}(A -\tilde{A})= 2(\bar\xi\xi + \bar\chi\chi)
\end{equation}
the generators
$I$, $\bar I$ form the $SU(1|1)$ algebra
$$
\{ I, \bar I \}_{{}_{\rm P}}= iB\,, \quad [ B, I ]_{{}_{\rm P}} = [ B, \bar I ]_{{}_{\rm P}}
= 0\,.
$$

The standard (odd) SUSY generators~(\ref{Q}), (\ref{S}) and
the even SUSY generators~(\ref{R}), (\ref{C})
form the $SU(1|1)$ supermultiplets:
the generators $I$, $\bar I$ transform
the even spinor generators~(\ref{R}), (\ref{C})
into the odd spinor ones~(\ref{Q}), (\ref{S}), and vice versa. We get
\be
\{I, Q_{\alpha} \}_{{}_{\rm P}}= R_{\alpha}\,, \; \{I, \bar S^{\dot\alpha} \}_{{}_{\rm P}}
= \bar C^{\dot\alpha} \,, \; [I, \bar R_{\dot\alpha} ]_{{}_{\rm P}}
= -\bar Q_{\dot\alpha}\,, \; [I, C^{\alpha} ]_{{}_{\rm P}}= - S^{\alpha} \quad \mbox{and c.~c.}\,.
\ee

The structure of our $SU(3,2|\,1)$ algebra can be depicted as follows
\be
\lefteqn{\underbrace{\phantom{R_\alpha\,\,\, \bar
R_{\dot\alpha}\,\,\, C^\alpha\,\,\, \bar C^{\dot\alpha}\,\,\, A \,\,\,\,\,\,\,
P_{\alpha\dot\beta}\,\,\, L_{\alpha\beta}\,\,\, \bar
L_{\dot\alpha\dot\beta}\,\,\, K^{\alpha\dot\beta}\,\,\,
D}}_{SU(3,2)}} R_\alpha\,\,\, \bar R_{\dot\alpha}\,\,\, C^\alpha \,\,\,
\bar C^{\dot\alpha}\,\,\, A \,\,\,\,\,\,\, \overbrace{P_{\alpha\dot\beta} \,\,\,
L_{\alpha\beta}\,\,\, \bar L_{\dot\alpha\dot\beta}\,\,\, K^{\alpha\dot\beta}\,\,\,
D \,\,\,\,\,\,\, Q_{\alpha}\,\,\, \bar
Q_{\dot\alpha}\,\,\, S^{\alpha}\,\,\,
\bar S^{\dot\alpha}\,\,\,\, \tilde{A}}^{SU(2,2|\,1)} \,\,\,\,\, I\,\,\,\,\, \bar I \,\,.
\ee
It is worth pointing out that the
odd SUSY generators $Q_{\alpha}$, $\bar Q_{\dot\alpha}$, $S^{\alpha}$,
$\bar S^{\dot\alpha}$ and odd $SU(1|1)$ generators $I$, $\bar I$,
reproduce the whole $SU(3,2\vert 1)$ superalgebra as their algebraic closure.

Using the canonical Poisson brackets~(\ref{cpb-tw}),
(\ref{CPB-tw}) and the expressions for the
generators~(\ref{P,K})--(\ref{A}), (\ref{Q})--(\ref{A2}), we find
how the supertwistor components transform under various
symmetries.
{} For example, the odd SUSY translations and
superconformal boosts act as
\begin{equation}\label{s-c-tw}
\delta \lambda_{\alpha} = 2i \eta_{\alpha}\chi
\, ,\qquad  \delta \bar\mu^{\dot\alpha} =
2i \bar\varepsilon^{\dot\alpha} \chi \,, \qquad
\delta \chi = \varepsilon^{\alpha}\lambda_{\alpha} -
\bar\eta_{\dot\alpha} \bar\mu^{\dot\alpha}\,,
\end{equation}
where $\varepsilon^\alpha$ and $\eta_\alpha$ are the relevant Grassmann parameters.
The transformations of the superspace variables of the system~(\ref{act-bsusy-f})
can be obtained from~(\ref{s-c-tw}) and the relations~(\ref{inc-f}), {\it e.g.}

\noindent
$\bullet$
odd SUSY translations
\begin{equation}\label{susy-f}
\delta x^{\dot\alpha\alpha} = i(\bar\varepsilon^{\dot\alpha}\theta^\alpha
-\bar\theta^{\dot\alpha}\varepsilon^\alpha) \, ,\qquad \delta
\theta^\alpha = \varepsilon^\alpha \, ,\qquad \delta
\zeta^{\alpha}= 0 \,;
\end{equation}
\noindent
$\bullet$
odd superconformal boosts
\begin{equation}\label{s-boost-f-x}
\delta x^{\dot\alpha\alpha} = i(\bar\theta^{\dot\alpha}\bar\eta_{\dot\beta}x^{\dot\beta\alpha}
- x^{\dot\alpha\beta}\eta_\beta \theta^\alpha )
+(\theta \eta +\bar\eta \bar\theta)\, \bar\theta^{\dot\alpha}\theta^\alpha
- (\bar\eta \bar\zeta)\, \bar\theta^{\dot\alpha}\zeta^\alpha
+ (\zeta \eta)\, \bar\zeta^{\dot\alpha}\theta^\alpha
 \, ,
\end{equation}
\begin{equation}\label{s-boost-f-t}
\delta \theta^\alpha = -2i (\theta \eta) \theta^\alpha
- \bar\eta_{\dot\beta} (x^{\dot\beta\alpha}+ i\bar\theta^{\dot\beta}\theta^\alpha
+ i\bar\zeta^{\dot\beta}\zeta^\alpha)\, ,\qquad
\delta \zeta^\alpha = - 2i\,(\zeta \eta) \, \theta^\alpha \, .
\end{equation}
Notice that the superconformal boosts mix the odd and even spinor coordinates.

The transformations produced by the generators~(\ref{I})
transform odd and even scalars into each other
\begin{equation}\label{I-c,k}
\delta \chi = \bar\sigma \xi \,,\qquad \delta \xi = - \sigma \chi\,.
\end{equation}
In terms of the superspace variables, the transformations~(\ref{I-c,k})
leads to the mixing of odd and even spinor coordinates
\begin{equation}\label{I-t,z}
\delta \theta^\alpha = \bar\sigma \zeta^\alpha \,,\qquad \delta \zeta^\alpha
= - \sigma \theta^\alpha\,.
\end{equation}

In the next Section we shall use the superspace
transformations~(\ref{s-boost-f-x}), (\ref{s-boost-f-t}), (\ref{I-t,z})
in order to find the $SU(3,2\vert 1)$ invariant form of the HS superfield equations.

\setcounter{equation}{0}
\section{Quantization}

\subsection{Irreducible first and second class constraints}

Now we shall perform the analysis of the constraints in the phase space formulation of
the model~(\ref{act-bsusy-f}).

We denote by $\pi_\alpha$ and $\bar \pi_{\dot\alpha}$
the momenta conjugate to $\zeta^\alpha$
and $\bar \zeta^{\dot\alpha}$, and by $P_\alpha$ and $\bar P_{\dot\alpha}$
the momenta conjugate to
$\theta^\alpha$ and $\bar \theta^{\dot\alpha}$.
The non-vanishing canonical Poisson brackets are
$$
[ x^{\dot\alpha\alpha} , P_{\beta\dot\beta}]_{{}_{\rm P}}
=\delta^{\alpha}_{\beta} \delta^{\dot\alpha}_{\dot\beta} \,,
\quad
\{ \theta^\alpha , P_{\beta} \} _{{}_{\rm P}} =\delta^{\alpha}_{\beta}  \,,
\quad \{ \bar\theta^{\dot\alpha} , \bar P_{\dot\beta}\}_{{}_{\rm P}}
=\delta^{\dot\alpha}_{\dot\beta} \,,
\quad
[ \zeta^\alpha , \pi_{\beta}]_{{}_{\rm P}} =\delta^{\alpha}_{\beta}
\,,\quad [ \bar \zeta^{\dot\alpha} , \bar\pi_{\dot\beta}]_{{}_{\rm P}}
=\delta^{\dot\alpha}_{\dot\beta} \,.
$$

The model~(\ref{act-bsusy-f}) is characterized by the mass-shell constraint
\begin{equation}\label{cons-P}
T \equiv P_{\alpha\dot\alpha}P^{\alpha\dot\alpha}\approx 0
\end{equation}
and two sets of spinor constraints. One set describes the odd ones
\begin{equation}\label{cons-Df}
D_\alpha\equiv P_\alpha
+iP_{\alpha\dot\alpha}\bar\theta^{\dot\alpha}\approx
0\,,\qquad\qquad \bar D_{\dot\alpha}\equiv \bar P_{\dot\alpha}
+i\theta^{\alpha}P_{\alpha\dot\alpha}\approx 0
\end{equation}
which are the same as in the Brink--Schwarz superparticle model~\cite{BS}. The second set
provides the even spinor constraints
\begin{equation}\label{cons-D}
{\nabla}_\alpha\equiv \pi_\alpha
+iP_{\alpha\dot\alpha}\bar\zeta^{\dot\alpha}\approx
0\,,\qquad\qquad \bar {\nabla}_{\dot\alpha}\equiv \bar\pi_{\dot\alpha}
-i\zeta^{\alpha}P_{\alpha\dot\alpha}\approx 0\,.
\end{equation}
These spinor constraints coincide with those in the HS bosonic particle model~\cite{FedLuk}.

Non-vanishing Poisson brackets of the
constraints~(\ref{cons-P})--(\ref{cons-D}) are
\begin{equation}\label{alg-D}
\{D_\alpha,\bar D_{\dot\alpha}\}_{{}_{\rm P}} =2iP_{\alpha\dot\alpha}\,,
\qquad
[ {\nabla}_\alpha,\bar {\nabla}_{\dot\alpha}]_{{}_{\rm P}} =2iP_{\alpha\dot\alpha}\,.
\end{equation}
The constraint~(\ref{cons-P}) is of first class. Due to the relation~(\ref{cons-P})
half of the spinor constraints,
both in the odd (see~(\ref{cons-Df})) and even sets (see~(\ref{cons-D})),
are of first class; the remaining constraints are of second class.
The first class constraints are singled out by multiplying
the spinor constraints by the spinor $\bar\zeta^{\dot\alpha}P_{\alpha\dot\alpha}$
and its complex conjugate, whereas the second class constraints are obtained by projecting
on $\zeta^{\alpha}$ and its conjugate. Thus the odd constraints~(\ref{cons-Df})
split into two second class constraints
\begin{equation}\label{f-2}
G\equiv \zeta^{\alpha} D_\alpha\approx 0\,,
\qquad \bar G\equiv \bar\zeta^{\dot\alpha}\bar D_{\dot\alpha}\approx 0
\end{equation}
and two first class constraints
\begin{equation}\label{f-1}
F\equiv \bar\zeta_{\dot\alpha}P^{\alpha\dot\alpha} D_\alpha\approx 0\,,
\qquad \bar F\equiv \zeta_{\alpha}P^{\alpha\dot\alpha}\bar D_{\dot\alpha}\approx 0\,.
\end{equation}
Analogously, the even constraints~(\ref{cons-D}) contain two second class constraints
\begin{equation}\label{b-2}
{\cal G}\equiv \zeta^{\alpha} {\nabla}_\alpha\approx 0\,, \qquad \bar {\cal G}
\equiv \bar\zeta^{\dot\alpha}\bar {\nabla}_{\dot\alpha}\approx 0
\end{equation}
and two first class constraints
\begin{equation}\label{b-1}
{\cal F}\equiv \bar\zeta_{\dot\alpha}P^{\alpha\dot\alpha} {\cal D}_\alpha\approx 0\,,
\qquad \bar {\cal F} \equiv \zeta_{\alpha}P^{\alpha\dot\alpha}\bar {\nabla}_{\dot\alpha}\approx 0\,.
\end{equation}
Poisson brackets of the constraints~(\ref{f-1}), (\ref{b-1})
are proportional to the mass-shell constraint~(\ref{cons-P})
whereas their brackets with the constraints~(\ref{f-2}), (\ref{b-2})
yield again~(\ref{f-1}), (\ref{b-1}).

The odd first class constraints~(\ref{f-1}) generate (irreducible)
local fermionic $\kappa$--symmetry. Analogously, the even first class constraints~(\ref{b-1})
generate a bosonic  $\kappa$--symmetry.

We shall subsequently apply the Gupta--Bleuler (GB) quantization
method \footnote{For the application of GB quantization method
to superparticle models see {\it e.~g.}~\cite{GB}.},
{\it i.~e.} we impose on wave
function all the first class constraints and half of the complex
second class constraints which (in weak sense, modulo constraints) commute. Remaining
complex--conjugated second class constraints are imposed on the
conjugated states.

Taking into consideration for example that the constraints $G\approx
0$ and $F \approx 0$ are equivalent to $D_\alpha \approx 0\,$,
etc., we can impose on the wave function $\Psi\,$, besides the constraint $T\,
\Psi = 0$, one set of the odd constraints:
\begin{equation}\label{ch-f}
\mbox{odd chiral set:}\qquad\qquad \bar D_{\dot\alpha}\, \Psi = 0\,, \qquad\quad
(\bar\zeta_{\dot\alpha}P^{\dot\alpha\alpha} D_\alpha)\, \Psi =  0\,,
\end{equation}
or
\begin{equation}\label{ach-f}
\mbox{odd anti-chiral set:}\qquad\qquad D_\alpha\, \Psi = 0\,,
\qquad\quad (\zeta_{\alpha}P^{\dot\alpha\alpha}\bar D_{\dot\alpha})\, \Psi = 0
\end{equation}
and one set of even constraints,
\begin{equation}\label{ch-b}
\mbox{even chiral set:}\qquad\qquad \bar {\nabla}_{\dot\alpha}\, \Psi = 0\,,
\qquad\quad  (\bar\zeta_{\dot\alpha}P^{\dot\alpha\alpha} {\nabla}_\alpha)\, \Psi = 0\,,
\end{equation}
or
\begin{equation}\label{ach-b}
\mbox{even anti-chiral set:}\qquad\qquad {\nabla}_\alpha\, \Psi = 0\,,
\qquad\quad (\zeta_{\alpha}P^{\dot\alpha\alpha}\bar {\nabla}_{\dot\alpha})\, \Psi = 0\,.
\end{equation}

\subsection{Superconformally invariant sets of the field equations}

We shall quantize the model~(\ref{act-bsusy-f}) by using the following
superwave function
\begin{equation}\label{wf}
\Psi = \Psi(x, \zeta, \bar\zeta, \theta, \bar\theta)
\end{equation}
consistent with the following realization of the momentum operators
$$
\hat P_{\alpha\dot\alpha}=-i\, \partial_{\alpha\dot\alpha}\,,
\;\,\,\, \hat\pi_{\alpha}=-i \,\partial/\partial \zeta^{\alpha}
\,, \; \,\,\, \hat{\bar\pi}_{\dot\alpha}=-i
\,\partial/\partial \bar\zeta^{\dot\alpha}\,,
\;\,\,\,
\hat P_{\alpha}= i \,\partial/\partial \theta^{\alpha} \,,
\quad\,\,\, \hat{\bar P}_{\dot\alpha}= i \, \partial/\partial
\bar \theta^{\dot\alpha}\,.
$$
Then the set of constraints defined
in~(\ref{cons-Df}), (\ref{cons-D}) become odd supercovariant
derivatives\footnote{We use the same notations for the covariant derivatives
and their classical counterparts~(\ref{cons-D}), (\ref{cons-Df}).}
\begin{equation}\label{op-Df}
D_\alpha= i(\partial/\partial \theta^{\alpha}
-i\partial_{\alpha\dot\alpha}\bar\theta^{\dot\alpha})\,, \qquad\qquad \bar D_{\dot\alpha}
= i(\partial/\partial \bar\theta^{\dot\alpha}
-i\theta^{\alpha}\partial_{\alpha\dot\alpha})
\end{equation}
and even covariant derivatives
\begin{equation}\label{op-D}
{\nabla}_\alpha= -i(\partial/\partial \zeta^{\alpha}
+i\partial_{\alpha\dot\alpha}\bar\zeta^{\dot\alpha})\,,\qquad\qquad
\bar {\nabla}_{\dot\alpha}= -i(\partial/\partial \bar\zeta^{\dot\alpha}
-i\zeta^{\alpha}\partial_{\alpha\dot\alpha})\,.
\end{equation}

The covariant derivatives~(\ref{op-Df}), (\ref{op-D}) and $\partial^{\dot\alpha\alpha}$
do transform under odd superconformal
bo\-osts~(\ref{s-boost-f-x}), (\ref{s-boost-f-t}) as follows
\begin{equation}\label{sc-D}
\delta D_\alpha= -2i \theta_{\alpha} \eta^{\beta} D_\beta
+ 2i (\bar\eta\bar\theta) D_{\alpha}+ 2i (\zeta\eta)  {\nabla}_{\alpha}\,,
\quad \delta \bar D_{\dot\alpha}= -2i \bar\theta_{\dot\alpha}
\bar\eta^{\dot\beta} \bar D_{\dot\beta}
- 2i (\theta\eta) \bar D_{\dot\alpha}+ 2i (\bar\eta\bar\zeta)  \bar {\nabla}_{\dot\alpha}\,,
\end{equation}
\begin{equation}\label{sc-N}
\delta {\nabla}_\alpha= 2i \eta_{\alpha} \theta^{\beta} {\nabla}_\beta
- 2i (\bar\eta\bar\zeta) D_{\alpha}\,, \qquad
\delta \bar {\nabla}_{\dot\alpha}= 2i \bar\eta_{\dot\alpha} \bar\theta^{\dot\beta}
\bar {\nabla}_{\dot\beta}
+ 2i (\zeta\eta) \bar D_{\dot\alpha}\,,
\end{equation}
\begin{equation}\label{sc-d}
\delta\, \partial^{\dot\alpha\alpha} = 2i  \partial^{\dot\alpha\beta} \theta_{\beta}\,
\eta^{\alpha} - 2i \bar\eta^{\dot\alpha} \,
\bar\theta_{\dot\beta} \partial^{\dot\beta\alpha}
+i {\bar D}^{\dot\alpha} \eta^{\alpha} - i \bar\eta^{\dot\alpha} D^{\alpha}\,.
\end{equation}
Thus the superconformal covariance requires $\Psi$ to
satisfy  either the chiral conditions~(\ref{ch-f}), (\ref{ch-b})
or the anti-chiral ones~(\ref{ach-f}), (\ref{ach-b}). We obtain two cases:

\noindent {\bf 1)} chiral case
\begin{equation}\label{c-c}
\Box\, \Psi= 0\,,\quad\,\,\,\,
\bar D_{\dot\alpha}\, \Psi= 0\,, \quad\,\,\,\, \bar {\nabla}_{\dot\alpha}\,
\Psi= 0\,, \quad\,\,\,\, (\bar\zeta_{\dot\alpha}\partial^{\dot\alpha\alpha} D_{\alpha})\,
\Psi =0\,, \quad\,\,\,\, (\bar\zeta_{\dot\alpha}\partial^{\dot\alpha\alpha}
{\nabla}_{\alpha})\,
\Psi =0\,;
\end{equation}
\noindent {\bf 2)} anti-chiral case
\begin{equation}\label{ac-ac}
\Box\, \Psi= 0\,,\quad\,\,\,\,
D_{\alpha}\, \Psi= 0\,, \quad\,\,\,\, {\nabla}_{\alpha}\, \Psi= 0\,,
\quad\,\,\,\, (\zeta_{\alpha}\partial^{\dot\alpha\alpha}\bar D_{\dot\alpha})\, \Psi =0\,,
\quad\,\,\,\, (\zeta_{\alpha}\partial^{\dot\alpha\alpha}\bar {\nabla}_{\dot\alpha})\, \Psi =0\,,
\end{equation}
where $\Box \equiv \partial^{\dot\alpha\alpha} \partial_{\alpha\dot\alpha}\,$.
We will consider further the chiral case~(\ref{c-c}), because the anti-chiral case can be
recovered by conjugation.

{}From the algebra of the spinor covariant derivatives~(\ref{op-Df}), (\ref{op-D})
\begin{equation}\label{alg-D-q}
\{D_\alpha,\bar D_{\dot\alpha}\} =2i \partial_{\alpha\dot\alpha} \,,
\qquad [ {\nabla}_\alpha,\bar {\nabla}_{\dot\alpha}] =2i \partial_{\alpha\dot\alpha}
\end{equation}
follows that the equation $\Box\, \Psi= 0$ is a consequence
of the other equations~(\ref{c-c}).
Also, by applying $\bar\nabla_{\dot\alpha}$ to the last two equations
in~(\ref{c-c}), it is easy to see that they amount to
\begin{equation}\label{Dir-cc}
\partial^{\dot\alpha\alpha} D_{\alpha}\, \Psi =0\,,
\qquad \partial^{\dot\alpha\alpha} {\nabla}_{\alpha}\, \Psi =0 \,.
\end{equation}
Thus, the set of the equations~(\ref{c-c}) is equivalent to the simpler set
\begin{equation}\label{c-c1}
\bar D_{\dot\alpha}\, \Psi= 0\,, \qquad \bar {\nabla}_{\dot\alpha}\, \Psi= 0\,,
\qquad \partial^{\dot\alpha\alpha} D_{\alpha}\, \Psi =0\,,
\qquad \partial^{\dot\alpha\alpha} {\nabla}_{\alpha}\, \Psi =0\,.
\end{equation}

Now we are prepared to check the superconformal invariance of these equations.
Using~(\ref{sc-D})--(\ref{sc-d}), we obtain the following variations
\begin{equation}\label{sc-b}
\delta\, \Box = -2i(\theta\eta-\bar\eta \bar\theta) \Box
-2i(\eta^{\alpha}\partial_{\alpha\dot\alpha} \bar D^{\dot\alpha}
+ \bar\eta_{\dot\alpha}\partial^{\dot\alpha\alpha} D_{\alpha})\, ,
\end{equation}
\begin{equation}\label{sc-dD}
\delta\, (\partial^{\dot\alpha\alpha} D_{\alpha})
= 4i(\bar\eta \bar\theta)\, \partial^{\dot\alpha\alpha} D_{\alpha}
+ 2i(\zeta \eta)\, \partial^{\dot\alpha\alpha} {\nabla}_{\alpha}
+ i\eta^{\alpha} D_{\alpha} \bar D^{\dot\alpha} - i\bar\eta^{\dot\alpha} D^{\alpha} D_{\alpha}
- \underline{ 2 \eta_{\alpha} \partial^{\dot\alpha\alpha}  } \, ,
\end{equation}
\begin{equation}\label{sc-dN}
\delta\, (\partial^{\dot\alpha\alpha} {\nabla}_{\alpha})
= - 2i(\theta \eta)\, \partial^{\dot\alpha\alpha} {\nabla}_{\alpha}
- 2i \bar\eta^{\dot\alpha} \bar\theta_{\dot\beta} \partial^{\dot\beta\beta} {\nabla}_{\beta}
- 2i(\bar\eta \bar\zeta)\, \partial^{\dot\alpha\alpha} D_{\alpha}
+ i\eta^{\alpha} {\nabla}_{\alpha} \bar D^{\dot\alpha}
- i\bar\eta^{\dot\alpha} D^{\alpha} {\nabla}_{\alpha}  \,,
\end{equation}
\begin{equation}\label{sc-DD}
\delta\, ({D}^{\alpha} {D}_{\alpha}) = 2i(\theta\eta+ 2\bar\eta \bar\theta) {D}^{\alpha}
{D}_{\alpha} - 4i(\zeta\eta) {D}^{\alpha} {\nabla}_{\alpha} -
\underline{ 4 \eta^{\alpha}{D}_{\alpha} } \,,
\end{equation}
\begin{equation}\label{sc-DN}
\delta\, ({D}^{\alpha} {\nabla}_{\alpha}) = 2i (\bar\eta \bar\theta)
{D}^{\alpha} {\nabla}_{\alpha} + 2i(\bar\eta \bar\zeta) {D}^{\alpha} {D}_{\alpha}
- \underline{ 2 \eta^{\alpha}{\nabla}_{\alpha} } \,.
\end{equation}
Having these transformation laws, one can draw two conclusions
concerning the superconformal covariance of the set \p{c-c1}. First,
the only way to cancel the underlined piece in \p{sc-dD} is to
assume for $\Psi$ the following transformation rule under the
conformal supersymmetry:
\begin{equation}\label{var-ch-f}
\delta\, \Psi = - 2i(\theta \eta)\, \Psi \,.
\end{equation}
Secondly, the superconformal covariance requires adding two
wave equations, besides those in \p{c-c1}, namely $D^{\alpha}
D_{\alpha}\, \Psi =0$ and $D^{\alpha} {\nabla}_{\alpha}\, \Psi =0$.
The underlined terms in the transformation rules \p{sc-DD},
\p{sc-DN} are then cancelled due to the transformation law
\p{var-ch-f} for $\Psi$, so the set of equations
\begin{equation}\label{eqs-cc}
\bar D_{\dot\alpha}\, \Psi=0 \,,\qquad \bar {\nabla}_{\dot\alpha}\,
\Psi=0 \,,\qquad D^{\alpha} D_{\alpha}\, \Psi =0 \,,\qquad
D^{\alpha} {\nabla}_{\alpha}\, \Psi =0
\end{equation}
is covariant under conformal supersymmetry. This set is also
covariant  under the $SU(1|1)$ odd transformations~(\ref{I-t,z})
\begin{equation}\label{su-D}
\delta D_\alpha= \sigma {\nabla}_{\alpha}\,,\qquad \delta \bar
D_{\dot\alpha} = - \bar\sigma \bar {\nabla}_{\dot\alpha}\,,\qquad
\delta {\nabla}_\alpha = \bar\sigma D_{\alpha}\,, \qquad \delta \bar
{\nabla}_{\dot\alpha}= \sigma  \bar D_{\dot\alpha}
\end{equation}
and consequently it is covariant under the closure of these transformations, {\it i.~e.}
the whole supergroup $SU(3,2\vert 1)\,$. In such a way we obtained the
minimal set \p{eqs-cc} of the $SU(3,2\vert 1)$ covariant equations. The
second order equations in \p{c-c1} now follow as the integrability
conditions of the set \p{eqs-cc}.

At this point we would like to mention an analogy with  the ordinary
$N=1$, $D=4$ superparticle. There, the GB quantization
with the wave superfunction $\Psi(x,\theta,\bar\theta)$
yields the first and third equations from the set \p{c-c1}
and they lead to the dynamical
equation $D^2\, \Psi(x,\theta) = a\,$ ($a$ is a complex number).
It is the requirement of
the superconformal $SU(2,2\vert 1)$ covariance that further sets the
constant $a$ equal to zero and produces the standard superconformally
invariant equation of motion for $N=1$, $D=4$ chiral superfield. In our
case the full set of eqs.~\p{c-c1} can be shown to specify
$D^2\,\Psi$ and $D\nabla\,\Psi$ up to two independent holomorphic
functions of $\zeta^\alpha$ comprising infinitely many constants
in their $\zeta$- expansions. The HS superconformal symmetry
$SU(3,2\vert 1)$ require that all these constants vanish, just as
$SU(2,2\vert 1)$ symmetry requires that $a=0$.
In both cases the requirement of superconformal
invariance can be regarded as an additional condition which allows
one to pass to physically more acceptable set
of the superfield dynamical equations. The presence of infinite number of
constants in our case is related to the fact that we are dealing
with an infinite number of $N=1$, $4D$ chiral superfields with growing
external helicities, what will become clear in the next Section.

Finally, it would be interesting to inquire whether the minimal
superconformal set of dynamical equations in \p{eqs-cc} can be
derived from some
modification of the
action principle in the extended
superspace\footnote{The off-shell superfield actions for massless
higher spins in the ordinary $N=1$, $D=4$ superspace were constructed
in \cite{K1,GK}.}.

\subsection{HS superconformal superwave function}

The chirality conditions
$\bar D_{\dot\alpha}\, \Psi=0$ and $\bar \nabla_{\dot\alpha}\, \Psi=0$ imply that
the superwave function $\Psi$ is defined on chiral superspace
\begin{equation}\label{cc-sp}
x_{\!{}_{L}}^{\dot\alpha\alpha} = x^{\dot\alpha\alpha} +i
\bar\zeta^{\dot\alpha} \zeta^{\alpha} +i \bar\theta^{\dot\alpha}
\theta^{\alpha} \,,\quad \zeta^{\alpha}  \,,\quad \theta^{\alpha}\,,
\end{equation}
i.e. $\Psi = \Psi_{\!{}_{L}}(x_{\!{}_{L}}, \zeta, \theta)\,$.  This
subspace is closed under superconformal transformations. In
particular, the odd SUSY translations and superconformal
boosts~(\ref{susy-f})--(\ref{s-boost-f-t}) act as
\begin{equation}\label{s-x}
\delta x_{\!{}_{L}}^{\dot\alpha\alpha} = 2i\bar\varepsilon^{\dot\alpha}\theta^\alpha -
2i x_{\!{}_{L}}^{\dot\alpha\beta} \eta_{\beta} \theta^{\alpha} \, , \qquad
\delta \theta^\alpha = \varepsilon^\alpha - 2i (\theta \eta) \, \theta^\alpha
- \bar\eta_{\dot\beta} x_{\!{}_{L}}^{\dot\beta\alpha}\, , \qquad
\delta \zeta^\alpha = - 2i (\zeta \eta)\, \theta^\alpha \, .
\end{equation}

The expansion of the superfield $\Psi_{\!{}_{L}}(x_{\!{}_{L}},
\zeta, \theta)$  with respect to the even spinor coordinate
\begin{equation}\label{wf-ch-1}
\Psi_{\!{}_{L}}
= \sum_{k=0}^{\infty}
\zeta^{\alpha_1}\ldots\zeta^{\alpha_k} \Phi^{(k)}_{\alpha_1
\ldots \alpha_k}(x_{\!{}_{L}}, \theta)
\end{equation}
contains chiral superfields with external symmetrized undotted spinor indices
\begin{equation}\label{ch-sf}
\Phi^{(k)}_{\alpha_1
\ldots \alpha_k}(x_{\!{}_{L}}, \theta) = A^{(k)}_{\alpha_1
\ldots \alpha_k} + \theta^\beta \psi^{(k+1)}_{(\alpha_1
\ldots \alpha_k\beta)} + \theta_{(\alpha_k} \varphi^{(k-1)}_{\alpha_1
\ldots \alpha_{k-1})}  + \theta^2 B^{(k)}_{\alpha_1
\ldots \alpha_k}\,.
\end{equation}
Eqs. $D^{\alpha} D_{\alpha}\, \Psi =0$ and $D^{\alpha} {\nabla}_{\alpha}\,
\Psi =0$ lead to Dirac equations for the first two component fields
\begin{equation}\label{Dir-comp}
\partial^{\dot\alpha\alpha_1} A^{(k)}_{\alpha_1
\ldots \alpha_k} =0\,, \qquad \partial^{\dot\alpha\alpha_1}
\psi^{(k+1)}_{\alpha_1 \ldots \alpha_{k+1}} =0
\end{equation}
and imply the vanishing of the last two components (which are thus auxiliary)
\begin{equation}\label{cons-B}
\varphi^{(k-1)}_{\alpha_1
\ldots \alpha_{k-1}} =0\,, \qquad B^{(k)}_{\alpha_1 \ldots \alpha_k}=0\,.
\end{equation}

Therefore, the space of states of our model is spanned by an infinite set of massless
supermultiplets~(\ref{ch-sf})
combined into a single HS chiral superwave function
$\Psi_{\!{}_{L}}(x_{\!{}_{L}}, \zeta, \theta)$
which satisfies the superconformally covariant equations~(\ref{eqs-cc}).
We point out that, like e.g. in \cite{BPST}, we are dealing with
the description of massless higher spins
in terms of superfield strengths, and eqs.~(\ref{eqs-cc})
encompass both the dynamical equations and Bianchi identities.

In our case due to the double expansion in $\zeta$ and $\theta$
each Grassmann-even helicity (both integer and half-integer)
has its Grassmann-odd ``partner''.
The same doubling of helicities initially takes place also in the case of the tensorial
HS superparticle quantized in twistor superspace (see {\it e.~g.}~\cite{BandLukSor}).
However, there one can avoid such a doubling by requiring the superwave function to have
a fixed parity under the reflection
%$\Lambda_a \rightarrow -\Lambda_a$
of the bosonic spinorial coordinates
and one can achieve the agreement
with spin--statistics theorem. In our case we can also impose similar
constraint with respect to the coordinates $\zeta^\alpha, \bar\zeta^{\dot\alpha}$,
but it will lead to the breaking of bosonic SUSY and $SU(3,2\vert 1)$ symmetry.
We would like to stress, however, that the problem of statistics
imposed on one--particle wave functions matters only if we
wish to consider the multiparticle states, or to perform second quantization.
We conclude therefore that the problem of the second quantization of
the superwave function~(\ref{wf-ch-1}) requires some additional
nonstandard input.

\subsection{Quantization in the twistor space and the twistor transform for HS superfields}

We shall present the quantization of pure twistorial model~(\ref{act-twist-s})
and establish its relation with the superspace description.

Let us consider the representation when the operators
of $\lambda_{\alpha}$, $\bar\mu^{\dot\alpha}$, $\chi$ and $\xi$ are diagonal.
The twistorial superwave function
$
{\tilde\Psi} = {\tilde\Psi}(\lambda, \bar\mu, \chi, \xi)
$
satisfies  the equation
\begin{equation}\label{H-q}
\left(\lambda_\alpha\frac{\partial}{\partial\lambda_\alpha}
+ {\bar\mu}^{\dot\alpha} \frac{\partial}{\partial{\bar\mu}^{\dot\alpha}}
+ \chi \frac{\partial}{\partial\chi} + \xi \frac{\partial}{\partial \xi}
\right) {\tilde\Psi}= c {\tilde\Psi}\,,
\end{equation}
which is a quantum version of the constraint~(\ref{H-s}).
The constant $c$ determining the homogeneity degree of the twistorial
superwave function takes care of the quantum ordering ambiguity.

Passing from the twistorial superwave function to
the chiral superfield $\Psi_{\!{}_{L}}(x_{\!{}_{L}}, \zeta, \theta)$ is accomplished by
the field twistor transform. For this purpose we make the substitutions~(\ref{inc-f})
$$
\bar \mu^{\dot\alpha}  = x_{\!{}_L}^{\dot\alpha\alpha}\lambda_\alpha  \,,\qquad
\chi = \theta^\alpha \lambda_\alpha  \,,\qquad
\xi = \zeta^\alpha \lambda_\alpha
$$
in the twistor superfield and use the integral transform
\begin{equation}\label{wf-tt}
\Psi_{\!{}_L} (x_{\!{}_L}, \theta, \zeta) = \oint (\lambda d\lambda)\,  \hat\Psi\!
\left(\lambda, \,x_{\!{}_L}\!\lambda , \,\theta \lambda , \,\zeta \lambda\right)
\end{equation}
where the contour integral is defined as in \cite{Fer}.
The twistorial superfield $\hat\Psi$ in the integral representation~(\ref{wf-tt}) has the
homogeneity degree $-2$ and is obtained by the following rescaling of $\tilde\Psi$
\begin{equation}\label{wf-tw}
\hat\Psi(\lambda, \bar\mu, \chi, \xi)=
\xi^{-c-2} {\tilde\Psi}(\lambda, \bar\mu, \chi, \xi)\,.
\end{equation}

Expanding the twistor superfield~(\ref{wf-tw}) in the following power series
\begin{equation}\label{wf-tw-1}
\hat\Psi\!\left(\lambda, \,x_{\!{}_L}\!\lambda , \,\theta \lambda , \,\zeta \lambda\right)
= \sum_{k=0}^{\infty} (\zeta \lambda)^k \hat\Phi{}^{(k)} (\lambda, \,x_{\!{}_L}\!\lambda ,
\,\theta \lambda)\,,
\end{equation}
and inserting~(\ref{wf-tw-1}) into~(\ref{wf-tt}), we obtain the standard twistor
transform for the chiral superfields~\cite{Fer}
\begin{equation}\label{ch-tt}
\Phi^{(k)}_{\alpha_1 \ldots \alpha_k}(x_{\!{}_{L}}, \theta)  = \oint (\lambda d\lambda)\,
\lambda_{\alpha_1}\ldots \lambda_{\alpha_k} \hat\Phi{}^{(k)}(\lambda, \,x_{\!{}_L}\!\lambda ,
\,\theta \lambda)\,,
\end{equation}
Note that the chiral superfields defined by the integral transformation~(\ref{ch-tt})
contain, if we expand them in powers of $\theta \lambda\,$,
only first two component fields because of the
identity $(\theta \lambda)^2\equiv 0$, i.e. they are automatically on shell
(see~(\ref{Dir-comp}), (\ref{cons-B})).

\setcounter{equation}{0}
\section{Summary and outlook}

We have proposed a new model for HS superparticle
propagating in $N=1$, $D=4$ superspace extended by commuting
Weyl spinor. We would like to point out that our HS superparticle
model possesses $SU(3,2|1)$ symmetry as an alternative to the $OSp(1|8)$ HS
generalization of conformal supersymmetry for tensorial superparticle.

The fundamental realization of the $OSp(1|8)$ conformal supersymmetry
is obtained by adding to real $Sp(8)$ spinors
one real odd component $\chi$ needed for the construction of $OSp(1|8)$ supertwistor.
This reality property of $\chi$ prevents us from
obtaining chiral supermultiplets in
the $OSp(1|8)$ HS theory. As we did show here, there is another way to
construct HS generalizations of superconformal group, achieved
by extending the conformal group $SU(2,2)$ to
$SU(3,2)\,$ in a way which allows one to preserve the important
notion of $N=1$ chirality\footnote{In the $OSp(1\vert 8)$
case the chirality can be introduced only at the cost of adding some extra
harmonic variables~\cite{IvLuk}.}.

We have shown that $SU(3,2)$ is the symmetry of
the massless particle models
with bosonic counterpart of SUSY.
As in the $Sp(8)$ case, the irreducible $SU(3,2)$ multiplets contain
the irreducible $SU(2,2)$ multiplets only once.
The further generalization to the supercase gives the $SU(3,2|1)$ symmetry
which unifies standard and ``bosonic'' conformal supersymmetries.
The bosonic coordinate sector containing second rank spinor
$x^{\dot\alpha\beta}$ and Weyl spinor $\zeta^\alpha$
is obtained via the following extension of the
conformal coset description of complex Minkowski space
(see {\it e.g.} \cite{We}):
\be
\frac{SU(2,2)}{S(U(2)\times U(2))} \qquad \longrightarrow \qquad
\frac{SU(3,2)}{S(U(3)\times U(2))}\,.
\ee
It is curious that in this way we gain the linearly realized
$SU(3)\times SU(2)\times U(1)$ symmetry.

Quantization of the HS superparticle was accomplished here by Gupta--Bleuler
method \cite{GB}. It should be pointed out
that for obtaining basic equations defining HS superconformal field
it was necessary
to exploit the superconformal invariance.
As a consequence, we get the
set of basic equations \p{eqs-cc} which is more restrictive than the set
of the equations~\p{c-c} following from the quantization of the
first class constraints.

In this paper we analyzed only the case of HS superparticle with the simple $N=1$
supersymmetries, both odd and even. Inclusion of the internal symmetry  group $SU(N)$
can be achieved by replacing the odd Grassmann component $\chi$
of $SU(3,2|1)$ supertwistor by
the odd $U(N)$ Grassmann spinor,
which leads to the extension
$\theta^\alpha  \rightarrow \theta^\alpha_i$, $i=1,...,N$
of the odd sector of superspace.
The corresponding superparticle model obtained by the generalization of
(\ref{act-bsusy-f}) will possess $SU(3,2|N)$ supersymmetry. There is,
however, another way to introduce the internal symmetry in HS theory.
One can use in the action~(\ref{act-bsusy}) $N$ even spinors $\zeta^\alpha_i$,
$i=1,...,N$. The symmetry of this model is $SU(2,2+N)$. In such a
way we also include the $U(N)$ internal symmetry
in the HS formalism. The most general model with the
$SU(2+N_1,2|N_2)$ conformal supersymmetry involving $N_1$ even Weyl
spinors $\zeta^\alpha_i$ ( $i=1,...,N_1$) and $N_2$ odd Weyl ones
$\theta^{\alpha}_k$ ($k=1,...,N_2$) can be
also considered.\footnote{In the twistorial framework
conformal and $N=1$ superconformal algebras extended by one or two even spinorial
generators were discussed in \cite {BP}.}

\section*{Acknowledgments}
S.F. would like to thank Bogoliubov--Infeld program for financial support and Wroc{\l}aw
University for warm hospitality. E.I. thanks Universitat de Valencia -- Departament de
Fisica Teorica and Laboratoire de Physique, ENS--Lyon, for kind hospitality at the
final stages of this study and J.A.~de Azc\'arraga, I.~Bandos,
D.~Sorokin for useful discussions. He also acknowledges
the partial support from the NATO grant PST.GLG.980302. The work of S.F. and E.I.
was partially supported by the RFBR grant 06-02-16684 and
a grant from Heisenberg-Landau program.

\end{document}